\documentstyle[eqsecnum,preprint,aps,prd]{revtex}
\preprint{WISC-MILW-94-TH-15}
\tighten
\begin{document}
\draft

\def\b{{\bf b}}
\def\a{{\bf a}}
\def\ut{{\tilde u}}
\def\vt{{\tilde v}}
\def\idu{{\int_0^1 du\>}}
\def\idv{{\int_0^1 dv\>}}
\def\idut{{\int_0^1 d\ut\>}}
\def\idvt{{\int_0^1 d\vt\>}}
\def\Ut{{\tilde U}}
\def\Vt{{\tilde V}}
\def\al{{\alpha}}
\def\E{{\dot E}}
\def\P{{\dot P}}
\def\apu{{{\bf a}'(u)}}
\def\bpv{{{\bf b}'(v)}}
\def\aput{{{\bf a}'(\ut)}}
\def\bpvt{{{\bf b}'(\vt)}}
\def\ccx{{\bpv \cdot \aput \times \bpvt}}
\def\ccy{{\apu \cdot \aput \times \bpvt}}
\def\ccz{{\apu \cdot \bpv \times \bpvt}}
\def\cct{{\apu \cdot \bpv \times \aput}}
\def\t{{\times}}


\title{CLOSED-FORM EXPRESSION FOR THE MOMENTUM RADIATED FROM COSMIC
STRING LOOPS}

\author{Bruce Allen}
\address{
Department of Physics, University of Wisconsin -- Milwaukee\\
P.O. Box 413, Milwaukee, Wisconsin 53201, U.S.A.\\
email: ballen@dirac.phys.uwm.edu}

\author{Paul Casper}
\address{
Department of Physics, University of Wisconsin -- Milwaukee\\
P.O. Box 413, Milwaukee, Wisconsin 53201, U.S.A.\\
email: pcasper@dirac.phys.uwm.edu}

\author{Adrian Ottewill}
\address{
Department of Mathematical Physics\\
University College Dublin, Belfield, Dublin 4, IRELAND\\
email: ottewill@relativity.ucd.ie}

\maketitle
\begin{abstract}
We modify the recent analytic formula given by Allen and Casper
\cite{AllenCasper} for the rate at which piecewise linear cosmic string
loops lose energy to gravitational radiation to yield the analogous
analytic formula for the rate at which loops radiate momentum.  The
resulting formula (which is exact when the effects of gravitational
back-reaction are neglected) is a sum of $O(N^4)$ polynomial and log
terms where, $N$ is the total number of segments on the piecewise linear
string loop. As illustration, we write the formula explicitly for a
simple one-parameter family of loops with $N=5$.  For most loops the
large number of terms makes evaluation ``by hand" impractical, but, a
computer or symbolic manipulator may by used to yield accurate
results.  The formula has been used to correct numerical results given
in the existing literature.  To assist future work in this area, a
small catalog of results for a number of simple string loops is
provided.
\end{abstract}


\section{INTRODUCTION}
\label{section1}
Cosmic strings are one-dimensional topological defects that may have
formed at phase transitions as the universe expanded and cooled
\cite{Kibble,Zel'dovich,Vilenkin,SV}.  A cosmic string loop is formed
when two sections of long string (a string with length greater than the
horizon length) meet and intercommute.  After a loop is formed, it
begins to oscillate under its own tension.  As cosmic string loops
oscillate, they lose energy in the form of gravitational radiation.
The formation and subsequent decay of cosmic string loops is of
fundamental importance to the evolution of the cosmic string network.
In addition, most of the observational limits on cosmic strings are
obtained by considering the effects of the gravitational radiation
emitted as the loops decay ($\!$\cite{SV,AllenCaldwell} and references
therein).

The power emitted in gravitational radiation by a cosmic string loop
depends upon its shape and velocity.  If the loop configuration is
asymmetric, then the energy may be radiated in an asymmetric way.  In
that case, the loop will radiate and lose momentum as well as energy.
In this paper, we obtain an analytic formula for the momentum radiated
for any piecewise linear cosmic string loop.

In the center-of-mass frame, a cosmic string loop is specified by the
position ${\bf x}(t,\sigma)$ of the string as a function of two
variables: time $t$ and a space-like parameter $\sigma$ that runs from
$0$ to $L$.  The total energy of the loop is $\mu L$ where $\mu$ is the
mass-per-unit-length of the string.  $L$ is referred to as the
``invariant length" of the loop.  When gravitational back-reaction is
neglected, the string loop satisfies equations of motion whose most
general solution in the center-of-mass frame is
\begin{equation}
{\bf x}(t,\sigma)= {1 \over 2} \big[ \a(t+\sigma) + \b(t
- \sigma)\big].
\label{X}
\end{equation}
Here $\a(u) \equiv \a(u+L)$ and $\b(v) \equiv \b(v+L)$ are a pair of
periodic functions, satisfying the ``gauge condition" $|\a'(u)| =
|\b'(v)|=1$, where $'$ denotes differentiation w.r.t. the function's
argument.  Because the functions $\a$ and $\b$ are periodic, each can
be described by a closed loop.  These loops are referred to respectively
as the $\a$-loop and the $\b$-loop.  Together, the $\a$- and $\b$-loops
define the trajectory of the string loop.

If we define the four-momentum of the gravity waves emitted by a string
loop to be $P^\al=(E,P^i)$ where $i=x,y,z$, then the average rate of
energy and momentum loss by an oscillating string loop is given by
the four-vector $-\P^\al$, where
\begin{equation}
\P^\al=(\E,\P^i) = \gamma^\al G \mu^2 .
\label{P0}
\end{equation}
Here $G$ is Newton's constant and we use units with $c=1$ and metric
signature $(-,+,+,+)$.  Throughout this paper, a dot appearing over a
symbol denotes the time derivative of that quantity.  In equation
(\ref{P0}), $\E$ is the energy radiated (i.e., the power) and $\P^i$
are the three spatial components of the momentum radiated, averaged
over a single oscillation of the loop.  With our definition of
$\gamma^\al$ and metric signature, the string loop is losing energy in
the form of gravitational radiation when $\gamma^0$ is positive (which
is always the case).  Note that in reference \cite{AllenCasper}, the
quantity $\gamma^0$ is denoted simply by $\gamma$.  When one of the
components of $\gamma^i$ is positive, the loop is radiating a net
amount of energy and momentum in that direction, and the loop itself
will recoil and begin to accelerate in the opposite direction.  Thus if
$\gamma^x >0$ then the loop will begin to accelerate in the $-x$
direction.  The dimensionless quantities
$\gamma^\al=(\gamma^0,\gamma^i)$ depend only upon the ``shape" of the
cosmic string loop.  That is, the energy and momentum radiated in
gravitational radiation from a loop is invariant under a rescaling
(magnification or shrinking) of the loop, provided that the velocity at
each point on the rescaled loop is unchanged
\cite{Vilenkin,AllenShellard}.  Thus, without loss of generality, we
consider only loops with invariant length $L=1$.

In a recent paper, we presented a new formula for $\gamma^0$.  The
formula is an exact analytic closed form for any piecewise linear
cosmic string loop \cite{AllenCasper}.  A piecewise linear loop is one
which, at any time, is composed of straight segments each of which has
constant velocity.  Equivalently, a piecewise linear loop is any loop
for which the corresponding $\a$- and $\b$-loops are piecewise linear.
As is shown in reference \cite{AllenCasper}, the piecewise linear
requirement is not very restrictive since in practice a smooth cosmic
string loop may be well approximated by a piecewise linear loop with a
moderately small number of segments $N$.

In the present paper, we show how the formula given in
\cite{AllenCasper} may be modified to give exact analytic closed form
for the spatial momentum, $\gamma^i$, as well as for $\gamma^0$.
The formula for the components of the momentum radiated are very
similar to the formula for the radiated power.  In each case, the
formulae is a sum of $O(N^4)$ terms, each of which involves nothing
more complicated than log or arctangent functions.  Our C-code, which
provides one implementation of the formulae, is publicly available via
anonymous FTP from the directory pub/pcasper at the internet site
alpha1.csd.uwm.edu.

The remainder of the paper is organized as follows.  Section
\ref{section2} explains how the formulae of reference
\cite{AllenCasper} may be modified to yield analytic, closed forms for
the three spatial components of the radiated momentum.  This section
generalizes the work done in section \ref{section3} of reference
\cite{AllenCasper}.  The final steps of the solution which are
described in sections IV-VI of reference
\cite{AllenCasper} are unchanged.  Thus, those sections are not
repeated in this paper.  In section \ref{section3} the resulting
formula for both the radiated power and momentum is written explicitly
for the case of a simple one-parameter family of string loops.  The
values of $\gamma^\al$ for this family of loops are compared to those
given by an independent numerical method as well as to the results
given by our C-code implementation of the general formulas.  Excellent
agreement is found in all cases.   The formula is then used to correct
the small number of numerical values for the momentum radiated which
appear in the existing literature.  These values are typically off by a
factor of 2, though in some cases they are off by as much as a factor
of 10. Section \ref{section4} contains a catalog of $\gamma^\al$ values
for some simple loop trajectories.  This is followed by a short
conclusion.


\section{MOMENTUM RADIATED BY COSMIC STRING LOOPS}
\label{section2}

In this section we derive a general formula for both the radiated
energy ($\E$) and the radiated momentum ($\P^i$) for an arbitrary
cosmic string loop.  As in reference \cite{AllenCasper}, we work in the
weak-field limit.  In this limit the gravitational back-reaction on the
loop in neglected so that the loop oscillates periodically in time.
This derivation is almost identical to the derivation of the formula
for the radiated energy given in section \ref{section3} of reference
\cite{AllenCasper}.  The difference is that the present derivation
yields the four-vector $\P^\al=(\E,\P^i)$.  The total radiated momentum
is simply given by $\P=(\P^i\P_i)^{1/2}$.  We find that the three
components of the radiated momentum may be obtained exactly for any
piecewise linear loop by the same method used previously for
the radiated energy.

The rate at which a loop loses four-momentum (averaged over a single
oscillation) is
\cite{VachVil,Durrer}
\begin{equation}
\P^\al=\gamma^\al G\mu^2=\sum_{n=0}^\infty\int d\Omega k^\al
{d\E_n\over d\Omega},
\label{P1}
\end{equation}
where the constant  four-vector $\gamma^\al=(\gamma^0,\gamma^i)$ depends
only upon the shape of the string loop's trajectory.  In equation (\ref
{P1}), the four-vector $k^\al$ is defined to be
$k^\al\equiv (1,{\bf \Omega})$ where
\begin{equation}
{\bf \Omega}\equiv (\sin\theta\cos\phi, \sin\theta\sin\phi, \cos\theta)
\label{Omega}
\end{equation}
is a unit spatial vector in the direction of the outgoing wave.
The integral $\int d\Omega$ appearing in (\ref{P1}) denotes integration
over angles on the two-sphere,
\begin{equation}
\int d\Omega\equiv \int_0^\pi\sin\theta d\theta\int_0^{2\pi}d\phi .
\label{twosphere}
\end{equation}
Because the loops oscillate with period $L/2=1/2$, they radiate only
at discrete angular frequencies given by
\begin{equation}
\omega_n=4\pi n\quad{\rm for}\quad n=1,2,3,\cdots.
\label{omega}
\end{equation}
The energy radiated per unit solid-angle into the $n$th mode
is given by
\begin{equation}
{d\E_n\over d\Omega}={G\over \pi}\omega_n^2[\tau_{\mu\nu}^*
(\omega_n{\bf \Omega})\tau^{\mu\nu}(\omega_n{\bf \Omega})-{1\over 2}|
\tau^\mu{}_\mu (\omega_n{\bf\Omega})|^2].
\label{dPdW}
\end{equation}
The Fourier transform of the stress tensor for a string
loop is
\begin{equation}
\tau_{\mu\nu}(\omega_n {\bf \Omega}) =2\mu \idu \idv G_{\mu\nu}(u,v)
e^{i \omega_n (u+v - {\bf \Omega} \cdot ({\bf a}(u) + {\bf b}(v)))/2},
\label{tau}
\end{equation}
where
\begin{equation}
G^{\mu\nu}(u,v)=\partial_ux^\mu\partial_vx^\nu+
\partial_vx^\mu\partial_ux^\nu,
\label{Guv}
\end{equation}
and $x^\mu=(t,{\bf x}(t,\sigma))$.  Combining equations (\ref{P1}),
(\ref{dPdW}) and (\ref{tau}), we find that
\begin{equation}
\P^\al= {2 G \mu^2 \over \pi} \sum_{n=-\infty}^{\infty}
 \omega_n^2 \int d\Omega
\idu \idv \idut \idvt
\psi(u,v,\ut,\vt) k^\al
e^{i \omega_n (\Delta t(u,v,\ut,\vt) - {\bf \Omega} \cdot \Delta
{\bf x}(u,v,\ut,\vt))}
\label{P2}
\end{equation}
where we have defined
\begin{eqnarray}
\psi(u,v,\ut,\vt) =&& G_{\mu\nu}(u,v) G^{\mu\nu}( \ut, \vt) -
{1 \over 2} G^\mu{}_\mu(u,v) G^\nu{}_\nu ( \ut ,\vt)\nonumber\\
=&&{1\over 8}[(\apu\cdot\aput -1)(\bpv\cdot\bpvt -1)+
(\apu\cdot\bpvt-1)(\bpv\cdot\aput -1)\nonumber\\
&&\qquad\qquad-(\apu\cdot\bpv -1)(\aput\cdot\bpvt -1)].
\label{psi}
\end{eqnarray}
The functions $\Delta t = (u+v- \ut-\vt)/2$ and $\Delta {\bf x}=({\bf
a}(u)+{\bf b}(v)-{\bf a}(\ut) - {\bf b}(\vt))/2$ in (\ref{P2}) describe
the temporal and spatial separation of the two points on the string
world-sheet with coordinates $(u,v)$ and $(\ut,\vt)$ respectively.  To
save space, in some of the formulae that follow, the arguments of
$\Delta t$ and $\Delta {\bf x}$ are not shown.  Since each term in the
sum over $n$ equals its complex-conjugate, as may be shown by
redefining $(u,v)\rightleftharpoons (\ut,\vt)$, $\P^\al$ is explicitly
real.  For this reason we have changed the sum over $n$ to a sum from
$-\infty$ to $\infty$ at the expense of introducing an overall factor
of $1/2$ into (\ref{P2}).  From here on, this sum will simply be
denoted by $\sum_n$.  Note that the timelike $\al=0$ component of
(\ref{P2}) is identical to the formula for the radiated energy given in
equation (3.8) of reference \cite{AllenCasper}.

It is possible to carry out both the sum over $n$ and the integral
over solid angle in (\ref{P2}) in closed form.
This is done by making use of the identity
\begin{equation}
\sum_n \omega_n \int d\Omega e^{i \omega_n (\Delta t - \Omega \cdot
\Delta {\bf x})} = 2 \pi i \sum_{k=-\infty}^{\infty} \epsilon(\Delta
t+k/2) \delta((\Delta t+k/2)^2 - |\Delta {\bf x}|^2),
\label{identity}
\end{equation}
where $\epsilon (x)=2\theta (x)-1$ is $+1$ for $x>0$ and $-1$ for
$x<0$ and $\delta$ is the Dirac delta function.  We define a
four-vector linear differential operator
\begin{equation}
D^\al(u,v,\ut,\vt) \equiv
U^\al(u,v,\ut,\vt) \partial_u+V^\al(u,v,\ut,\vt) \partial_v -
\Ut^\al (u,v,\ut,\vt) \partial_{\ut} -
\Vt^\al (u,v,\ut,\vt) \partial_{\vt},
\label{Ddef}
\end{equation}
where the vector functions $U^\al,V^\al,\Ut^\al$, and $\Vt^\al$ are
defined by the effect of $D^\al$ on the exponential:
\begin{equation}
D^\al \exp ({i \omega_n [\Delta t - {\bf \Omega} \cdot \Delta {\bf x}]})=i
\omega_n k^\al \exp ({i \omega_n [\Delta t - {\bf \Omega} \cdot \Delta
{\bf x}]}).
\label{Dexp}
\end{equation}
Because $D^\al$ is chosen to be a linear differential operator,
(\ref{Dexp}) is equivalent to the 16 equations
\begin{eqnarray}
D^0 \Delta t(u,v,\ut,\vt) = 1, \qquad \qquad D^0 \Delta x_j
(u,v,\ut,\vt) = 0,\nonumber\\
D^i \Delta t(u,v,\ut,\vt) = {\bf 0}, \qquad \qquad D^i \Delta x_j
(u,v,\ut,\vt) = -\delta^i{}_j,
\label{Ddef2}
\end{eqnarray}
where $\delta^i{}_j$ is the Kronecker delta function.  With this
definition of the operator $D^\al$, the identity
(\ref{identity}) can be used to express the radiated energy and
momentum (\ref{P2}) as
\begin{eqnarray}
\P^\al=&& 4 G \mu^2 \sum_{k=-\infty}^{\infty} \idu \idv \idut \idvt
\psi(u,v,\ut,\vt)\nonumber\\
&&\qquad\qquad\qquad D^\al(u,v,\ut,\vt) \bigg[\epsilon( \Delta t + k/2)
\delta( (\Delta t + k/2)^2-|\Delta {\bf x}|^2)\bigg].
\label{P3}
\end{eqnarray}
This equation has the same functional form as equation (3.15) of reference
\cite{AllenCasper}.  By the exact same arguments as those given
following (3.15) in \cite{AllenCasper}, we find that equation
(\ref{P3}) may be written
\begin{equation}
\P^\al= 8 G \mu^2 \idu \idv \idut \int_{-2}^2 d\vt \>
\psi(u,v,\ut,\vt) D^\al \bigg[\theta( \Delta t)
\delta( (\Delta t)^2-|\Delta {\bf x}|^2)\bigg].
\label{Pfinal}
\end{equation}
All that remains is to write down the explicit expression for the
differential operator $D^\al$.

The operator $D^\al$ is determined by the equations
given in (\ref{Ddef2}).
These equations may be written in matrix form as
\begin{equation}
\left(\matrix{1&1&1&1\cr \cr
- {\bf a}'(u) & -{\bf b}'(v)& -{\bf a}'(\tilde u)&-{\bf b}'(\tilde v) \cr}
\right)
\left(\matrix{U^\al\cr V^\al \cr{\tilde U}^\al \cr{\tilde V}^\al\cr}\right)
               = 2 I
\label{mateq}
\end{equation}
where $I$ denotes the identity matrix. Equation (\ref{mateq})
may be solved to find $U^\al$,
$V^\al$, $\Ut^\al$,  and $\Vt^\al $ (and therefore $D^\al$)
explicitly. If we denote minus twice the inverse determinant of the
prefactor matrix by
\begin{eqnarray}
Q(u,v,\ut,\vt)=2\bigg(\ccx&&-\ccy+\nonumber\\
&&\ccz-\cct\bigg)^{-1},
\label{Q}
\end{eqnarray}
then we find that the coefficients of the partial derivatives in $D^0$
are
\begin{eqnarray}
U^0&=& \ccx\> Q, \qquad V^0=- \ccy\> Q, \nonumber\\
\Ut^0&=&\ccz\> Q, \qquad \Vt^0= - \cct\> Q, \nonumber\\
\label{D0}
\end{eqnarray}
which is exactly the result found in reference \cite{AllenCasper}.  In
addition, the coefficients which define the operators $D^i$
are given by
\begin{eqnarray}
U^i&=&\bigg(\bpv \t \aput +\bpvt \t \bpv +\aput \t \bpvt\bigg)^i
\>Q,\nonumber\\
V^i&=&\bigg(\bpv \t \apu +\apu \t \bpvt +\bpvt \t \aput\bigg)^i
\>Q,\nonumber\\
\Ut^i&=&\bigg(\apu \t \bpv +\bpv \t \bpvt +\bpvt \t \apu\bigg)^i
\>Q,\nonumber\\
\Vt^i&=&\bigg(\bpv \t \apu +\apu \t \aput +\aput \t \bpv\bigg)^i\>Q.
\label{Di}
\end{eqnarray}
Equation (\ref{Pfinal}) may now be expressed in closed form for any
piecewise linear loop by exactly the same method as given in sections IV-VI
of \cite{AllenCasper}.  The {\it only} difference between the
calculation of the radiated energy and the calculation of a component of
the momentum radiated is in the choice of the coefficients of the
differential operator $D^\al$.


\section{RESULTS}
\label{section3}

The formula (\ref{Pfinal}) for $\P^\al$ may be evaluated exactly for
any piecewise linear cosmic string loop.  The general solution is given
in sections IV-VI of reference \cite{AllenCasper}.  In most cases the
large number of terms involved makes it impractical to write out the
solution explicitly, however, there are cases where the solution may
be written in a manageable form.  As an example, in section
\ref{threeA} we give the closed form solution for a one-parameter
family of cosmic string loops obtained with the aid of {\sl Mathematica}.
In the cases where it is impractical to
write out the closed-form solution, accurate values of $\gamma^\al$ may
still be obtained using a computer implementation of the general
formula.  In section \ref{threeB} one such implementation is used to
correct the small number of numerical values for the radiated momentum
which appear in the existing literature.

\subsection{Analytic Results}
\label{threeA}

We will now give closed forms for $\P^\al$ for a one-parameter family
of string loops.  These are defined by $\a$- and $\b$-loops consisting
of 2 and 3 segments respectively. The $\a$-loop is taken to lie along
the $z$-axis.  One kink on the $\a$-loop is positioned at the origin;
the parameter $u=0$ at this kink.  The other kink (at $u=1/2$) is
positioned above the first kink and has coordinates $(0,0,1/2)$.  The
three-segment $\b$-loop has the shape of an equilateral triangle.
For the $\b$-loop, we again position one kink at the origin and set
the parameter $v=0$ at that kink.  The position of the other two kinks
depends on a parameter $\phi$.  The kink at $v=1/3$ has coordinates $-
{1\over 6} (\cos\phi,\sqrt{3},\sin\phi)$ and the kink at $v=2/3$ has
coordinates ${1\over 6}(\cos\phi,-\sqrt{3},\sin\phi)$. If we make the
definition $s\equiv\sin\phi$, then the rate of energy loss by this set
of string loops may be written as
\begin{eqnarray}
\E(s)&=&{16G\mu^2\over(1-s^2)(4-s^2)^2}\times\nonumber\\
&&\>\bigg(12(1 - s^2)(4 + s^2)\log(2) + 18(4 - 3s^2)\log(3) -
(1 - s^2)(2 - s)^2(1 + 2s)\log(1 - s) \nonumber\\
&&\quad -(1 - s^2)(2 + s)^2(1 - 2s)\log(1 + s)-
(1 - s)(4 - s^2)^2\log(2 - s) \nonumber\\
&&\qquad-(1 + s)(4 - s^2)^2\log(2 + s) -
(1 - s^2)(2 - s)^2(4 - s)\log(4 - s) \nonumber\\
&&\qquad\quad -(1 - s^2)(2 + s)^2(4 + s)\log(4 + s)\bigg).
\label{angam}
\end{eqnarray}
The $x$ and $y$ components of the momentum radiated by these loops both
vanish.  The $z$ component is given by
\begin{eqnarray}
\P^z(s)&=&{16G\mu^2\over (1-s^2)(4-s^2)^2}\times\nonumber\\
&&\quad\bigg( -48s(1-s^2)\log(2) + 18s^3 \log(3) -
(1-s^2)(2-s)^2(1+2s) \log(1-s)\nonumber\\
&&\qquad+(1-s^2)(2+s)^2(1-2s) \log(1+s)+
(1-s)(4-s^2)^2 \log(2-s) \nonumber\\ &&\qquad\quad-
(1+s)(4-s^2)^2 \log(2+s)
-(1-s^2)(2-s)^2(4-s) \log(4-s) \nonumber\\ &&\qquad\qquad+
(1-s^2)(2+s)^2(4+s) \log(4+s)\bigg).
\label{pangam}
\end{eqnarray}
Note that $\E(s)$ and $\P(s)$ are even and odd functions of $s$
respectively, as they must be.
The exact results given by (\ref{angam}) and (\ref{pangam}) have been
compared to those given by our C-code implementation of the general
formula as well as to the independent numerical results given by the
Fast Fourier Transform method of Allen and Shellard
\cite{AllenShellard}.  The different sets of results are compared in
Figure 1.  We find excellent agreement between all three sets of
results.

\subsection{Corrected Results}
\label{threeB}

In this section we compare the results for the radiated momentum given
by other authors in the previous literature to the values given by our
formula.  The only family of loop trajectories for which numerical
values of the radiated momentum have been previously published is a
two-parameter family of loops first studied by Vachaspati and Vilenkin
\cite{VachVil}.  The $\a$- and $\b$-loops which define these
trajectories are given by
\begin{eqnarray}
{\bf a}(u)&=&{1\over 2\pi}[\sin(2\pi u){\bf \hat x} -\cos(2\pi u)
(\cos\phi{\bf \hat y} +\sin\phi{\bf \hat z})]\nonumber\\
{\bf b}(v)&=&{1\over 2\pi}\bigl[\bigl({\alpha\over 3}
\sin(6 \pi v)-(1-\alpha)\sin(2\pi v)\bigr){\bf \hat x} \nonumber\\
&&\quad\quad-\bigl({\alpha\over 3}\cos(6 \pi v)+(1-\alpha)
\cos(2\pi v)\bigr){\bf \hat y} \nonumber\\
&&\quad\quad\quad\quad\quad-(\alpha(1-\alpha))^{1/2}
\sin(4\pi v){\bf \hat z}\bigr]
\label{VV}
\end{eqnarray}
where $\alpha$ and $\phi$ are constant parameters, $0\le\alpha\le 1$
and $-\pi <\phi\le\pi$.  These loops have also been studied by Durrer
\cite{Durrer}.  In both cases, the authors determined $\P^\al$ by
numerically evaluating a finite number of terms of the infinite sum
appearing in (\ref{P1}), and then adding an estimate of the
contribution from the truncated infinite ``tail".  It is shown in
reference \cite{AllenCasper} that the errors in the numerical values of
$\gamma^0$ calculated by this method are typically due to errors in the
estimates of the contribution from the tail.  This appears to also be
the case for the values of $\P^i$ calculated by this method, which are
typically off by a factor of 2.  The results found by Vachaspati and
Vilenkin, and Durrer, are shown in Figures 2 and 3 along with the
results of our new method for the cases $\al=0.5$ and $\al=0.8$.  The
results given by our formula are found by evaluating $\gamma^\al$ for a
piecewise linear loop with approximately the same shape as the smooth
loops given by (\ref{VV}).  The approximation becomes more accurate as
the number of piecewise linear segments used is increased.  The
piecewise linear loops whose $\gamma^\al$ values where used in Figures
2 and 3 had 100 segments for both the $\a$- and $\b$-loops.  With this
number of segments, the error due to the piecewise linear approximation
is estimated to be no more than 5 percent.


\section{CATALOG OF LOOPS}
\label{section4}

This section gives a short catalog of piecewise linear loop
trajectories and their $\gamma^\al$ values.  This catalog expands the
catalog given in section VIII of reference \cite{AllenCasper} by including
both the radiated energy and the three components of the radiated
momentum for a number of simple loop trajectories.  The values given in
this catalog are intended (as in the original catalog) to be ``bench
mark" values which might prove useful in testing future analytic or
numerical methods.  In fact, the original catalog has already proven to
be quite useful.  The trajectories chosen for this catalog are the same
five two-parameter families of loops used in the original catalog.
The values of $\gamma^\al$ are shown as column vectors of the form
\begin{equation}
\gamma^\al=\left(\matrix{\gamma^0\cr \gamma^x\cr
\gamma^y\cr \gamma^z \cr}\right).
\label{colvector}
\end{equation}
For the readers convenience, a full description of the $\a$- and
$\b$-loops which define these trajectories is repeated here.

The first set of trajectories we consider are defined by $\a$- and
$\b$-loops consisting of 2 and 3 segments respectively.    The
$\a$-loop is taken to lie along the $z$-axis.  One kink on the $\a$-loop
is positioned at the origin; the parameter $u=0$ at this kink.  The
other kink (at $u=1/2$) is positioned above the first kink and has
coordinates $(0,0,1/2)$. The three segment $\b$-loop has the shape of
an equilateral triangle.  We again position one kink at the origin and
set the parameter $v=0$ at that kink.  The position of the other two
kinks depends on two parameters $\phi$ and $\theta$.  When
$\phi=\theta=0$, the $\b$-loop lies in the $x$-$z$ plane.  The kink at
$v=1/3$ has coordinates $(-1/6,0,\sqrt{3}/6)$ and the kink at $v=2/3$
has coordinates $(1/6,0,\sqrt{3}/6)$.  When $\phi$ and $\theta$ are not
zero, the position of the $\b$-loop is found as follows.  First, the
$\b$-loop is rotated by the angle $\phi >0$ about the $z$-axis
(counter-clockwise when viewed from large positive $z$).  After the
$\phi$ rotation, the $\b$-loop is then rotated by the angle $\theta >0$
about the $x$-axis (counter-clockwise when viewed from large positive
$x$).  Values of $\gamma^\al$ for the trajectories defined by these
$\a$- and $\b$-loops are given in Table \ref{table1} for several values
of the angles $\phi$ and $\theta$. It should be noted that the
one-parameter family of loops obtained by setting $\theta=90$ degrees
is equivalent to the family of loops for which the analytic formulas
(\ref{angam}) and (\ref{pangam}) are given in section \ref{section3}.

The second set of trajectories we consider are defined by $\a$- and
$\b$-loops consisting of 3 segments each.  Both the $\a$- and
$\b$-loops are equilateral triangles.  The position of the $\b$-loop
depends on the two parameters $\phi$ and $\theta$ in exactly the same
way as the $\b$-loop in the first set of trajectories.  The $\a$-loop
is placed in the same position the $\b$-loop has for parameter values
$\phi=\theta=0$.  Values of $\gamma^\al$ for the trajectories defined
by these $\a$- and $\b$-loops are given in Table \ref{table2} for
several values of the angles $\phi$ and $\theta$.

The third set of trajectories we consider are defined by $\a$- and
$\b$-loops consisting of 2 and 5 segments respectively.  The two
segment $\a$-loop is identical to the $\a$-loop used in the first set
of trajectories.  The $\b$-loop is taken to be a pentagon.  One kink on
the pentagon is positioned at the origin and is chosen to have
parameter value $v=0$.  When $\phi=\theta=0$, the $\b$-loop lies in the
$x$-$z$ plane, and is positioned so that the kink at $v=1/5$ has
coordinates $-{1\over 5} (\cos(\pi/5),0,-\sin(\pi/5))$, the kink at
$v=2/5$ has coordinates ${1\over 5}((\sin(\pi/10)-\cos(\pi/5)),0,
(\cos(\pi/10)+\sin(\pi/5)))$, and so on.  When $\phi$ and $\theta$ are
not equal to zero, the $\b$-loop is rotated in exactly the same manner
as for the previous two sets of loop trajectories.   Values of
$\gamma^\al$ for the trajectories defined by these $\a$- and $\b$-loops
are given in Table \ref{table3} for several values of the angles $\phi$
and $\theta$.

The fourth set of trajectories we consider are defined by $\a$- and
$\b$-loops consisting of 5 and 3 segments respectively.  The $\a$-loop
is a pentagon placed in the same position as the $\b$-loop in the third
trajectory set for parameter values $\phi=\theta=0$.  The $\b$-loop is
an equilateral triangle whose position is given in terms of the
parameters $\phi$ and $\theta$ in exactly the same way as the
$\b$-loops used in the first and second trajectory sets.  Values of
$\gamma^\al$ for the trajectories defined by these $\a$- and $\b$-loops
are given in Table \ref{table4} for several values of the angles $\phi$
and $\theta$.

The final set of trajectories we consider are defined by $\a$- and
$\b$-loops consisting of 5 segments each.  Both loops are taken to be
pentagons.  The $\a$-loop is identical to the $\a$-loop used in the
fourth set of trajectories.  The $\b$-loop is in the same position as
the $\a$-loop when $\phi=\theta=0$.  When $\phi$ and $\theta$ are not
zero, the $\b$-loop is rotated in the same manner as in the previous
sets of trajectories.  Values of $\gamma^\al$ for the trajectories
defined by these $\a$- and $\b$-loops are given in Table \ref{table5}
for several values of the angles $\phi$ and $\theta$.

The simple loop trajectories and their $\gamma^\al$ values given in
this section should provide a convenient reference against which any
future numerical or analytical methods may be tested.  Additional tests
are provided by comparison to the analytic results given in section
\ref{section3} and to the large number of simple analytical results
(for $\E$) given in a recent paper by the authors \cite{ACO}.  Finally,
comparisons may be made, for any loop, to the results given by the
computer implementation of our general formula.


\section{CONCLUSION}
\label{conclusion}

We have modified the method of Allen and Casper \cite{AllenCasper} to
yield analytic closed-form results for the linear momentum radiated by
piecewise linear cosmic string loops.  Any cosmic string loop can be
arbitrarily well approximated by a piecewise linear loop with the
number of segments sufficiently large.  An exact formula is given for a
simple one-parameter family of string loops.  Our computer
implementation of the general formula is then used to investigate the
small number of numerical results published in the previous
literature.  These results are found to be typically off by a factor of
2 from the correct results, though in some cases they are off by as
much as a factor of 10.  A small catalog of loop trajectories and their
$\gamma^\al$ values has been provided as a set of bench mark results
for future analytic or numerical work.  Although the string loops
studied in this paper are not physically realistic, they provide a
simple set of trajectories with which to test our formula.  We intend
to use the method of this paper to investigate a large sample of
more physically realistic loop trajectories in the near future.


\noindent
\vskip 0.3in
\centerline{ACKNOWLEDGMENTS}
\vskip 0.05in
This work was supported in part by NSF Grant No. PHY-91-05935 and a NATO
Collaborative Research Grant.



\begin{figure}
\caption{The power ($\E$) and radiated momentum ($\P$) for a simple one
parameter family of loop trajectories given in section
\protect\ref{threeA}.  The solid lines are the exact results given in
(\protect\ref{angam}) and (\protect\ref{pangam}).  The crosses are the
results given by the C-code implementation of the general formula and
the triangles are the results of the numerical Fast Fourier Transform
method.  There is excellent agreement between all three sets of
results.}
\label{fig1}
\end{figure}

\begin{figure}
\caption{Values of $\P=(\P^i\P_i)^{1/2}$ are shown (in units of
$G\mu^2$) for the loop trajectories given in (\protect\ref{VV}) with
$\al=0.5$.  The results given by our formula are shown by the solid
line.  Durrer's results are shown as dots while the results of
Vachaspati and Vilenkin are shown as crosses.}
\label{fig2}
\end{figure}

\begin{figure}
\caption{Values of $\P=(\P^i\P_i)^{1/2}$ are shown (in units of
$G\mu^2$) for the loop trajectories given in (\protect\ref{VV}) with
$\al=0.8$.  The results given by our formula are shown by the solid
line.  Durrer's results are shown as dots.}
\label{fig3}
\end{figure}


\begin{table}
\caption{The $\gamma^\al$ values (shown as column vectors) for the
first two-parameter family of string loops described in section
\protect\ref{section4}.}
\begin{tabular}{ccccc}
$\theta/\phi$&$18^\circ$&$36^\circ$&$54^\circ$&$72^\circ$\\
\tableline
&&&&\\
$18^\circ$
&$\left (\matrix{\hfill 59.80\cr\hfill 0.00\cr\hfill 0.00\cr\hfill
-1.90\cr}\right )$
&$\left (\matrix{\hfill 61.21\cr\hfill 0.00\cr\hfill 0.00\cr\hfill
-4.05\cr}\right )$
&$\left (\matrix{\hfill 63.51\cr\hfill 0.00\cr\hfill 0.00\cr\hfill
-6.62\cr}\right )$
&$\left (\matrix{\hfill 66.30\cr\hfill 0.00\cr\hfill 0.00\cr\hfill
-9.45\cr}\right )$\\
&&&&\\
$36^\circ$
&$\left (\matrix{\hfill 54.56\cr\hfill 0.00\cr\hfill 0.00\cr\hfill
-1.53\cr}\right )$
&$\left (\matrix{\hfill 56.86\cr\hfill 0.00\cr\hfill 0.00\cr\hfill
-3.47\cr}\right )$
&$\left (\matrix{\hfill 60.93\cr\hfill 0.00\cr\hfill 0.00\cr\hfill
-6.46\cr}\right )$
&$\left (\matrix{\hfill 67.45\cr\hfill 0.00\cr\hfill 0.00\cr\hfill
-11.79\cr}\right )$\\
&&&&\\
$54^\circ$
&$\left (\matrix{\hfill 50.15\cr\hfill 0.00\cr\hfill 0.00\cr\hfill
-0.67\cr}\right )$
&$\left (\matrix{\hfill 52.56\cr\hfill 0.00\cr\hfill 0.00\cr\hfill
-1.37\cr}\right )$
&$\left (\matrix{\hfill 56.40\cr\hfill 0.00\cr\hfill 0.00\cr\hfill
-2.09\cr}\right )$
&$\left (\matrix{\hfill 60.72\cr\hfill 0.00\cr\hfill 0.00\cr\hfill
-2.74\cr}\right )$\\
&&&&\\
$72^\circ$
&$\left (\matrix{\hfill 47.54\cr\hfill 0.00\cr\hfill 0.00\cr\hfill
-0.12\cr}\right )$
&$\left (\matrix{\hfill 50.12\cr\hfill 0.00\cr\hfill 0.00\cr\hfill
0.04\cr}\right )$
&$\left (\matrix{\hfill 54.47\cr\hfill 0.00\cr\hfill 0.00\cr\hfill
1.02\cr}\right )$
&$\left (\matrix{\hfill 60.36\cr\hfill 0.00\cr\hfill 0.00\cr\hfill
3.70\cr}\right )$\\
\end{tabular}
\label{table1}
\end{table}

\begin{table}
\caption{The $\gamma^\al$ values (shown as column vectors) for the
second two-parameter family of string loops described in section
\protect\ref{section4}.}
\begin{tabular}{ccccc}
$\theta/\phi$&$18^\circ$&$36^\circ$&$54^\circ$&$72^\circ$\\
\tableline
&&&&\\
$18^\circ$
&$\left (\matrix{\hfill 100.85\cr\hfill -2.93\cr\hfill 0.88\cr\hfill
-8.77\cr}\right )$
&$\left (\matrix{\hfill 90.65\cr\hfill -6.34\cr\hfill -0.34\cr\hfill
-12.02\cr}\right )$
&$\left (\matrix{\hfill 74.48\cr\hfill -3.17\cr\hfill -0.81\cr\hfill
-6.40\cr}\right )$
&$\left (\matrix{\hfill 64.80\cr\hfill -1.55\cr\hfill -0.64\cr\hfill
-3.72\cr}\right )$\\
&&&&\\
$36^\circ$
&$\left (\matrix{\hfill 82.00\cr\hfill 3.06\cr\hfill 1.17\cr\hfill
-2.01\cr}\right )$
&$\left (\matrix{\hfill 76.01\cr\hfill -1.79\cr\hfill 0.88\cr\hfill
-5.24\cr}\right )$
&$\left (\matrix{\hfill 70.97\cr\hfill -2.46\cr\hfill 0.27\cr\hfill
-6.31\cr}\right )$
&$\left (\matrix{\hfill 64.82\cr\hfill -1.42\cr\hfill -0.03\cr\hfill
-4.58\cr}\right )$\\
&&&&\\
$54^\circ$
&$\left (\matrix{\hfill 72.61\cr\hfill 5.16\cr\hfill 0.93\cr\hfill
0.06\cr}\right )$
&$\left (\matrix{\hfill 66.51\cr\hfill 0.44\cr\hfill 0.83\cr\hfill
-1.45\cr}\right )$
&$\left (\matrix{\hfill 63.68\cr\hfill -0.77\cr\hfill 0.54\cr\hfill
-2.56\cr}\right )$
&$\left (\matrix{\hfill 61.54\cr\hfill -0.64\cr\hfill 0.33\cr\hfill
-2.28\cr}\right )$\\
&&&&\\
$72^\circ$
&$\left (\matrix{\hfill 67.04\cr\hfill 6.10\cr\hfill 0.49\cr\hfill
0.91\cr}\right )$
&$\left (\matrix{\hfill 61.24\cr\hfill 1.38\cr\hfill 0.46\cr\hfill
0.08\cr}\right )$
&$\left (\matrix{\hfill 59.12\cr\hfill -0.06\cr\hfill 0.38\cr\hfill
-0.59\cr}\right )$
&$\left (\matrix{\hfill 58.80\cr\hfill -0.26\cr\hfill 0.30\cr\hfill
-0.27\cr}\right )$\\
\end{tabular}
\label{table2}
\end{table}

\begin{table}
\caption{The $\gamma^\al$ values (shown as column vectors) for the
third two-parameter family of string loops described in section
\protect\ref{section4}.}
\begin{tabular}{ccccc}
$\theta/\phi$&$18^\circ$&$36^\circ$&$54^\circ$&$72^\circ$\\
\tableline
&&&&\\
$18^\circ$
&$\left (\matrix{\hfill 64.15\cr\hfill 0.00\cr\hfill 0.00\cr\hfill
1.00\cr}\right )$
&$\left (\matrix{\hfill 66.04\cr\hfill 0.00\cr\hfill 0.00\cr\hfill
2.30\cr}\right )$
&$\left (\matrix{\hfill 69.28\cr\hfill 0.00\cr\hfill 0.00\cr\hfill
4.37\cr}\right )$
&$\left (\matrix{\hfill 74.52\cr\hfill 0.00\cr\hfill 0.00\cr\hfill
8.45\cr}\right )$\\
&&&&\\
$36^\circ$
&$\left (\matrix{\hfill 54.99\cr\hfill 0.00\cr\hfill 0.00\cr\hfill
0.27\cr}\right )$
&$\left (\matrix{\hfill 57.78\cr\hfill 0.00\cr\hfill 0.00\cr\hfill
0.54\cr}\right )$
&$\left (\matrix{\hfill 62.31\cr\hfill 0.00\cr\hfill 0.00\cr\hfill
0.70\cr}\right )$
&$\left (\matrix{\hfill 67.63\cr\hfill 0.00\cr\hfill 0.00\cr\hfill
0.44\cr}\right )$\\
&&&&\\
$54^\circ$
&$\left (\matrix{\hfill 48.74\cr\hfill 0.00\cr\hfill 0.00\cr\hfill
0.03\cr}\right )$
&$\left (\matrix{\hfill 52.02\cr\hfill 0.00\cr\hfill 0.00\cr\hfill
-0.03\cr}\right )$
&$\left (\matrix{\hfill 57.78\cr\hfill 0.00\cr\hfill 0.00\cr\hfill
-0.53\cr}\right )$
&$\left (\matrix{\hfill 66.36\cr\hfill 0.00\cr\hfill 0.00\cr\hfill
-2.94\cr}\right )$\\
&&&&\\
$72^\circ$
&$\left (\matrix{\hfill 45.30\cr\hfill 0.00\cr\hfill 0.00\cr\hfill
0.00\cr}\right )$
&$\left (\matrix{\hfill 48.74\cr\hfill 0.00\cr\hfill 0.00\cr\hfill
-0.03\cr}\right )$
&$\left (\matrix{\hfill 54.98\cr\hfill 0.00\cr\hfill 0.00\cr\hfill
-0.10\cr}\right )$
&$\left (\matrix{\hfill 64.04\cr\hfill 0.00\cr\hfill 0.00\cr\hfill
-0.16\cr}\right )$\\
\end{tabular}
\label{table3}
\end{table}

\begin{table}
\caption{The $\gamma^\al$ values (shown as column vectors) for the
fourth two-parameter family of string loops described in section
\protect\ref{section4}.}
\begin{tabular}{ccccc}
$\theta/\phi$&$18^\circ$&$36^\circ$&$54^\circ$&$72^\circ$\\
\tableline
&&&&\\
$18^\circ$
&$\left (\matrix{\hfill 84.69\cr\hfill 1.98\cr\hfill 0.76\cr\hfill
-1.27\cr}\right )$
&$\left (\matrix{\hfill 75.43\cr\hfill -0.84\cr\hfill 0.13\cr\hfill
-2.31\cr}\right )$
&$\left (\matrix{\hfill 67.82\cr\hfill -1.21\cr\hfill -0.14\cr\hfill
-2.88\cr}\right )$
&$\left (\matrix{\hfill 62.44\cr\hfill -0.80\cr\hfill -0.29\cr\hfill
-3.47\cr}\right )$\\
&&&&\\
$36^\circ$
&$\left (\matrix{\hfill 77.37\cr\hfill 3.40\cr\hfill 0.97\cr\hfill
-0.52\cr}\right )$
&$\left (\matrix{\hfill 71.03\cr\hfill 0.28\cr\hfill 0.39\cr\hfill
-1.97\cr}\right )$
&$\left (\matrix{\hfill 65.71\cr\hfill -0.55\cr\hfill 0.23\cr\hfill
-2.93\cr}\right )$
&$\left (\matrix{\hfill 61.86\cr\hfill -0.63\cr\hfill 0.28\cr\hfill
-3.92\cr}\right )$\\
&&&&\\
$54^\circ$
&$\left (\matrix{\hfill 70.11\cr\hfill 4.76\cr\hfill 0.77\cr\hfill
0.93\cr}\right )$
&$\left (\matrix{\hfill 65.69\cr\hfill 1.68\cr\hfill 0.54\cr\hfill
0.12\cr}\right )$
&$\left (\matrix{\hfill 62.72\cr\hfill 0.69\cr\hfill 0.72\cr\hfill
-0.38\cr}\right )$
&$\left (\matrix{\hfill 60.42\cr\hfill 0.06\cr\hfill 0.82\cr\hfill
-0.84\cr}\right )$\\
&&&&\\
$72^\circ$
&$\left (\matrix{\hfill 64.41\cr\hfill 5.50\cr\hfill 0.42\cr\hfill
1.81\cr}\right )$
&$\left (\matrix{\hfill 61.37\cr\hfill 2.74\cr\hfill 0.26\cr\hfill
1.74\cr}\right )$
&$\left (\matrix{\hfill 61.13\cr\hfill 2.35\cr\hfill 0.65\cr\hfill
2.93\cr}\right )$
&$\left (\matrix{\hfill 61.33\cr\hfill 1.41\cr\hfill 1.25\cr\hfill
4.46\cr}\right )$\\
\end{tabular}
\label{table4}
\end{table}

\begin{table}
\caption{The $\gamma^\al$ values (shown as column vectors) for the
fifth two-parameter family of string loops described in section
\protect\ref{section4}.}
\begin{tabular}{ccccc}
$\theta/\phi$&$18^\circ$&$36^\circ$&$54^\circ$&$72^\circ$\\
\tableline
&&&&\\
$18^\circ$
&$\left (\matrix{\hfill 114.46\cr\hfill -4.06\cr\hfill -0.32\cr\hfill
-2.21\cr}\right )$
&$\left (\matrix{\hfill 94.04\cr\hfill 0.72\cr\hfill -0.32\cr\hfill
4.07\cr}\right )$
&$\left (\matrix{\hfill 80.22\cr\hfill 1.21\cr\hfill 0.03\cr\hfill
5.06\cr}\right )$
&$\left (\matrix{\hfill 68.84\cr\hfill 0.61\cr\hfill 0.08\cr\hfill
3.32\cr}\right )$\\
&&&&\\
$36^\circ$
&$\left (\matrix{\hfill 93.52\cr\hfill -2.28\cr\hfill 0.37\cr\hfill
-2.47\cr}\right )$
&$\left (\matrix{\hfill 82.49\cr\hfill -1.88\cr\hfill -0.32\cr\hfill
-1.40\cr}\right )$
&$\left (\matrix{\hfill 72.06\cr\hfill -0.11\cr\hfill -0.20\cr\hfill
0.70\cr}\right )$
&$\left (\matrix{\hfill 65.40\cr\hfill 0.14\cr\hfill -0.09\cr\hfill
1.04\cr}\right )$\\
&&&&\\
$54^\circ$
&$\left (\matrix{\hfill 77.15\cr\hfill 0.47\cr\hfill 0.36\cr\hfill
-0.55\cr}\right )$
&$\left (\matrix{\hfill 72.94\cr\hfill -1.48\cr\hfill 0.03\cr\hfill
-1.56\cr}\right )$
&$\left (\matrix{\hfill 67.22\cr\hfill -0.34\cr\hfill -0.10\cr\hfill
-0.44\cr}\right )$
&$\left (\matrix{\hfill 62.74\cr\hfill 0.04\cr\hfill 0.02\cr\hfill
-0.06\cr}\right )$\\
&&&&\\
$72^\circ$
&$\left (\matrix{\hfill 67.11\cr\hfill 1.50\cr\hfill 0.15\cr\hfill
0.18\cr}\right )$
&$\left (\matrix{\hfill 64.76\cr\hfill -0.40\cr\hfill 0.11\cr\hfill
-0.41\cr}\right )$
&$\left (\matrix{\hfill 64.05\cr\hfill 0.30\cr\hfill 0.06\cr\hfill
0.55\cr}\right )$
&$\left (\matrix{\hfill 62.98\cr\hfill 0.58\cr\hfill 0.26\cr\hfill
1.89\cr}\right )$\\
\end{tabular}
\label{table5}
\end{table}

\end{document}